# Spark plasma sintering synthesis of ReB$_2$-type medium-entropy diboride (W$_{1/3}$Re$_{1/3}$Ru$_{1/3}$)B$_2$ with high hardness

Wuzhang Yang[1,2,3], Guorui Xiao[1,2,4], Zhi Ren[1,2,*]

[1]*School of Science, Westlake University, 18 Shilongshan Road, Hangzhou 310024, P. R. China*

[2]*Institute of Natural Sciences, Westlake Institute for Advanced Study, 18 Shilongshan Road, Hangzhou 310024, P. R. China*

[3]*Department of Physics, Fudan University, Shanghai 200433, P. R. China*

[4]*Department of Physics, Zhejiang University, Hangzhou 310027, P. R. China*

\* Corresponding author.

E-mail: renzhi@westlake.edu.cn

**Abstract:** A new medium-entropy diboride (MEDB) (W$_{1/3}$Re$_{1/3}$Ru$_{1/3}$)B$_2$ has been synthesized by spark plasma sintering of elemental powders at 1600 ºC. Despite the dissimilar structures of WB$_2$, ReB$_2$ and RuB$_2$, the sintered MEDB consists of a single hexagonal ReB$_2$-type phase (space group *P*6$_3$/*mmc*) with a relative density of 94.2% and an average grain size of 6.8 ± 2.2 μm. Structural refinement and electron microscopy measurements show that the W, Re, and Ru atoms occupy the same crystallographic site and are distributed uniformly in the lattice. The (W$_{1/3}$Re$_{1/3}$Ru$_{1/3}$)B$_2$ MEDB has Vickers hardnesses of 30.7 GPa at a load of 0.49 N and 20.5 GPa at a load of 9.8 N, which are comparable or higher than those reported for individual binary counterparts.

Transition metal diborides (TMDBs) have been studied extensively over the past few decades due to their excellent mechanical [1], anticorrosion [2], catalytic [3] and physical properties [4], which can be exploited for diverse applications. Recently, the development of TMDBs based on the entropy concept, including medium-entropy diborides (MEDBs) and high-entropy diborides (HEDBs), has gained increasing attention [5–18]. Examples include $(Ti_{0.2}Zr_{0.2}Hf_{0.2}Nb_{0.2}Ta_{0.2})B_2$, $(Ti_{0.2}Zr_{0.2}Hf_{0.2}Mo_{0.2}Ta_{0.2})B_2$, $(Ti_{0.2}Zr_{0.2}Hf_{0.2}Nb_{0.2}Mo_{0.2})B_2$, $(Ti_{0.2}Zr_{0.2}Hf_{0.2}Mo_{0.2}W_{0.2})B_2$ and (Zr-Nb-Ta)$B_2$. All these diborides adopt the hexagonal AlB$_2$-type structure (space group $P6/mmm$) [19], which is built up by two-dimensional transition metal and boron layers stacked alternatively along the $c$-axis [Fig. 1(a)]. In each transition metal layer, three or more different kinds of transition metal atoms are randomly distributed at the same crystallographic site, which results in an enhanced configurational entropy. As a consequence, the MEDBs and HEDBs often exhibit superior hardness and oxidation resistance compared with the average of individual binary counterparts [5, 10, 14].

Apart from the AlB$_2$-type structure, TMDBs can also exist in other structural types depending on the arrangement of boron atoms. In fact, WB$_2$ possesses both planar and puckered boron layers [Fig. 1(b)], leading to transformation of the space group into $P6_3/mmc$ [20]. While ReB$_2$ shares the same space group with WB$_2$, the former consists solely of the puckered boron layers [Fig. 1(c)] and is known to be an ultra-incompressible material [21]. As for RuB$_2$, it has an orthorhombic structure (space group $Pmmn$) and does not contain boron layers at all [Fig. 1(d)] [22]. To our knowledge, however, HEDBs and MEDBs with structure other than the AlB$_2$-type have not been explored to date.

Here we present the spark plasma sintering (SPS) synthesis of the new $(W_{1/3}Re_{1/3}Ru_{1/3})B_2$ MEDB. X-ray diffraction and electron microcopy investigations indicate that a single, highly crystalline ReB$_2$-type phase is obtained and all the constituent elements are distributed uniformly on both a macroscopic and microscopic scale. The microhardness of the MEDB is measured as a function of applied load up to 9.8 N and compared with related systems. The implications of these results are also discussed.

High purity W (99.99%), Re (99.99%), Ru (99.99%) and B (99.99%) powders with a stoichiometric ratio of W:Re:Ru:B = 1/3:1/3:1/3:2 were mixed thoroughly in an argon-filled glovebox. The mixed powders with a total mass of 2 g were then loaded into the graphite die with 10 mm inner diameter and sintered in a SPS apparatus at 1600°C in argon atmosphere under 20 MPa. The heating rate was 100°C/min and the holding time was 25 min. Given the short sintering time, the carbon diffusion from graphite dies only affects the surface layer of the sample, which was removed by mechanical polishing. After treatment, the sample has a disk shape with a height of 2.2 mm and its photograph is shown in Fig. S1 of the Supplemetary Material. For comparison, binary diborides $WB_2$, $ReB_2$ and $RuB_2$ were also prepared by the arc melting method. The crystal structure of resulting samples was examined by powder x-ray diffraction (XRD) using a Bruker D8 Advance x-ray diffractometer with the Cu-$K\alpha$ radiation. The morphology and elemental distribution of the MEDB were investigated by a Zeiss Supratm 55 Schottky field emission scanning electron microscope (SEM) equipped with an energy dispersive x-ray (EDX) spectrometer. The microstructure was examined in an FEI Tecnai G2 F20 S-TWIN transmission electron microscope (TEM) operated under an accelerating voltage of 200 kV. The Vickers micro-hardness $H_V$ of the MEDB was measured using an automatic microhardness testing system with applied loads $P$ varying from 0.49 N to 9.8 N, and $H_V$ is given as

$$H_V = 1854.4 P/r^2, \qquad (1)$$

where $r$ is the mean length of the two indention diagonals. For each $P$, the data are averaged from three independent measurements and the dwelling time for each measurement is 10s. It is pointed out that the averaged values from three measurements are almost the same as those from four or more measurements and hence good enough in terms of statistical representation.

Figure 1(e) shows the XRD patterns for the $(W_{1/3}Re_{1/3}Ru_{1/3})B_2$ MEDB as well as binary diborides $WB_2$, $ReB_2$, and $RuB_2$. The peak indexing indicates that all the samples are of a single phase except for $WB_2$, where a small amount of $WB_4$ is also present. The lattice parameters determined by Le-Bail fitting using the JANA2006 programme are $a$ = 2.9870 Å and $c$ = 13.8932 Å for $WB_2$, $a$ = 2.9009 Å and $c$ = 7.4868 Å for $ReB_2$, and $a$ = 4.6437 Å, $b$ = 2.8666 Å, and $c$ = 4.0455 Å for $RuB_2$. These values agree well with those reported previously [20–22]. Notably, the diffraction pattern for $(W_{1/3}Re_{1/3}Ru_{1/3})B_2$ are very similar to that of $ReB_2$ but distinct from those of $WB_2$ and $RuB_2$, indicating that the former two have essentially the same crystal structure. Indeed, as shown in Fig. 1(f), all the diffraction peaks can be well fitted based on the $ReB_2$- type structure with the $P6_3/mmc$ space group and no carbide impurity is detected. Here it is assumed that the W, Re and Ru atoms occupy the same (0.3333, 0.6667, 0.25) site and the boron atoms occupy the (0.3333, 0.6667, 0.548) site. While the refinement of the atomic position is not allowed in the Le-Bail method, the calculated XRD pattern agrees well with the observed one with reliability factors $R_{wp}$ = 9.9% and $R_p$ = 7.0%, confirming the validity of this structural model. From the refinement, one obtains $a$ = 2.9050 Å and $c$ = 7.5122 Å, which are slightly larger than those of $ReB_2$. This lattice expansion is attributed to the incorporation of W (1.40 Å), which has a lager atomic radius than both Re (1.38Å) and Ru (1.35 Å) [23].

Figure 2(a) shows the SEM image with a scale bar of 20 $\mu$m for the fracture surface of a broken $(W_{1/3}Re_{1/3}Ru_{1/3})B_2$ MEDB. One can see that the sample is compact and the grains are of an elongated shape. The latter is expected to contribute to the XRD peak broadening, which has already been taken into consideration in the structural refinement. After smoothing the surface with sandpaper, a few pores with size less than 1 $\mu$m are observed, as seen in Fig. 2(b). Nonetheless, the EDX elemental maps show that the W, Re, Ru, and B elements are distributed uniformly [see. Fig. 2(c-f)]. In addition, the average W:Re:Ru ratio is determined to be 0.97(3):0.94(6):1.09(9), in good agreement with the nominal one within the experimental error. The electron back-scatter diffraction (EBSD) phase and orientation maps for the MEDB are displayed in Fig. 2(g) and (h), respectively. The results confirm that the MEDB consists of a single $ReB_2$-type phase with random grain orientations. The grain size distribution is shown in Fig. 2(i), which gives an average grain size of 6.8 ± 2.2 $\mu$m. In addition, the relative density is found to be 94.2% for the (W$_{1/3}$Re$_{1/3}$Ru$_{1/3}$)B$_2$ MEDB. These parameters are listed in Table 1.

The results of TEM characterization of the (W$_{1/3}$Re$_{1/3}$Ru$_{1/3}$)B$_2$ MEDB are summarized in Fig. 3. As seen from Fig. 3(a), the high-resolution TEM (HRTEM) image taken along the [0 0 1] zone axis reveals a triangular lattice with a spacing of 0.239 nm, which matches well with the (100) plane. This is corroborated by the selected-area electron diffraction (SAED) shown in Fig. 3(b). Indeed, a well-defined spot pattern is observed and the spots near the center are indexable to (100), (010) and (110) planes. The HRTEM image and SAED pattern taken along the [1 2 0] zone axis are displayed in Figs. 3(c) and (d), respectively. In the HRTEM image, clear lattice fringes with a spacing of 0.374 nm are resolved, which corresponds well with the (001) plane. In addition, the spots near the center of SAED pattern can be indexed to (002), (2-10) and (2-12) planes. Figure 3(e) shows the TEM image of the MEDB with a scale bar of 250 nm. The EDX elemental mapping of the area marked by the frame ascertains the uniform distribution of all the constituent elements. The overall results affirm that the (W$_{1/3}$Re$_{1/3}$Ru$_{1/3}$)B$_2$ MEDB is highly crystalline and homogeneous down to nanometer scale.

The measured $H_V$ of the (W$_{1/3}$Re$_{1/3}$Ru$_{1/3}$)B$_2$ MEDB is plotted as a function of the applied load $P$ in Fig. 3 and typical indentation images are shown in Fig. S2 of the Supplemetary Material. For comparison, the literature data for binary WB$_2$ [20], ReB$_2$ [21] and RuB$_2$ [22] are also included. At a load of $P$ = 9.8 N, the $H_V$ of the (W$_{1/3}$Re$_{1/3}$Ru$_{1/3}$)B$_2$ MEDB is determined to be 20.5 ± 0.8 GPa. With decreasing $P$, the $H_V$ increases rapidly and reaches 30.7 ± 2.4 GPa at 0.49 N. This value is close to that of ReB$_2$ (31.3 ± 2.3 GPa at 0.49 N) [20] but ~25% and ~50% larger than those of WB$_2$ (24.4 ± 0.7 GPa at 0.49 N) [21] and RuB$_2$ (20.6 ± 1.9 GPa at 0.49 N) [22], respectively. In fact, for $P$ > 0.49 N, the $H_V$-$P$ curve of the MEDB lies above all those of the binary diborides. While the RuB$_2$ sample used in Ref. [22] was prepared by the arc-melting method and its microstructure was not characterized, the $H_V$ data of both WB$_2$ [20] and ReB$_2$ [21] were collected on samples prepared by SPS with relative densities of ~82% and ~99%, respectively. Hence the microhardness of the MEDB appears to be enhanced compared with the binary counterparts, which is presumably due to the solid solution effect. Nonetheless, it is prudent to note that the $H_V$ of this ReB$_2$-type MEDB is lower than those of AlB$_2$-type HEBs containing WB$_2$ [14, 25], suggesting that the former may be further improved by increasing the relative density and refining the microstructure.

Finally, we briefly discuss the implications of our results. First, our results demonstrate that a single phase MEDB can be formed even if all the individual binary counterparts are of different structures. This unusual phase stability is reminiscent of that observed in the high-entropy silicide $(V_{1/5}Cr_{1/5}Nb_{1/5}Ta_{1/5}W_{1/5})_5Si_3$ [24] and opens up more possibilities of transition metal combinations to compose the MEDBs and HEDBs of different structural types. Second, it is noted that the valence electron concentration of the $(W_{1/3}Re_{1/3}Ru_{1/3})B_2$ MEDB is the same as ReB$_2$ but higher than those of the AlB$_2$-type HEDBs. Hence, one may suspect that the valence electron concentration plays a role in determining the structure of MEDBs and HEDBs. In this regard, tracing the structural evolution in compositionally varied $(W_xRe_yRu_z)B_2$ MEDBs is of interest for future studies.

In summary, we have synthesized a new MEDB, $(W_{1/3}Re_{1/3}Ru_{1/3})B_2$, by the SPS method. The MEDB adopts a single ReB$_2$-type structure, in which the W, Re and Ru atoms share the same crystallographic site. Electron microscopy characterizations indicate that all the constituent elements are distributed uniformly on both a macroscopic and microscopic scale. The $H_V$ of the $(W_{1/3}Re_{1/3}Ru_{1/3})B_2$ MEDB is 20.5 GPa at 9.8 N and increased to 30.7 GPa at 0.49 N. These values are comparable or higher than those of binary WB$_2$, ReB$_2$ and RuB$_2$. Our study expands the structural diversity of MEDBs and calls for further exploration of similar multicomponent diborides, which may facilitate their design and application.

**Acknowledgements**

We thank the foundation of Westlake University for financial support and the Service Center for Physical Sciences for technical assistance in SEM measurements.

**Declaration of competing interest**

The authors have no competing interests to declare that are relevant to the content of this article.

**Table 1.** Structural parameters, relative density, grain size and Vickers hardness for the $(W_{1/3}Re_{1/3}Ru_{1/3})B_2$ MEDB. The available data for binary $WB_2$, $ReB_2$ and $RuB_2$ are also included for comparison.

| Diboride | Space group | Lattice parameters | Relative density | Grain size (μm) | $H_V$ (GPa) |
|---|---|---|---|---|---|
| $(W_{1/3}Re_{1/3}Ru_{1/3})B_2$ | $P6_3/mmc$ | $a$ = 2.9050 Å, $c$ = 7.5122 Å | 94.2% | 6.8 ± 2.2 | 30.7 ± 2.4[a]<br>20.5 ± 0.8[b] |
| $WB_2$ | $P6_3/mmc$ | $a$ = 2.9870 Å, $c$ = 13.8932 Å | ~82% | – | 24.4 ± 0.7[c] |
| $ReB_2$ | $P6_3/mmc$ | $a$ = 2.9009 Å, $c$ = 7.4868 Å | ~99% | – | 31.2 ± 2.3[d] |
| $RuB_2$ | $Pmmn$ | $a$ = 4.4637 Å, $b$ = 2.8666 Å, $c$ = 4.4055 Å | – | – | 20.6 ± 1.9[e] |

[a] This work at 0.49 N
[b] This work at 9.8 N
[c] Ref. [20] at 0.49 N
[d] Ref. [21] at 0.49 N
[e] Ref. [22] at 0.49 N

**Figure 1**

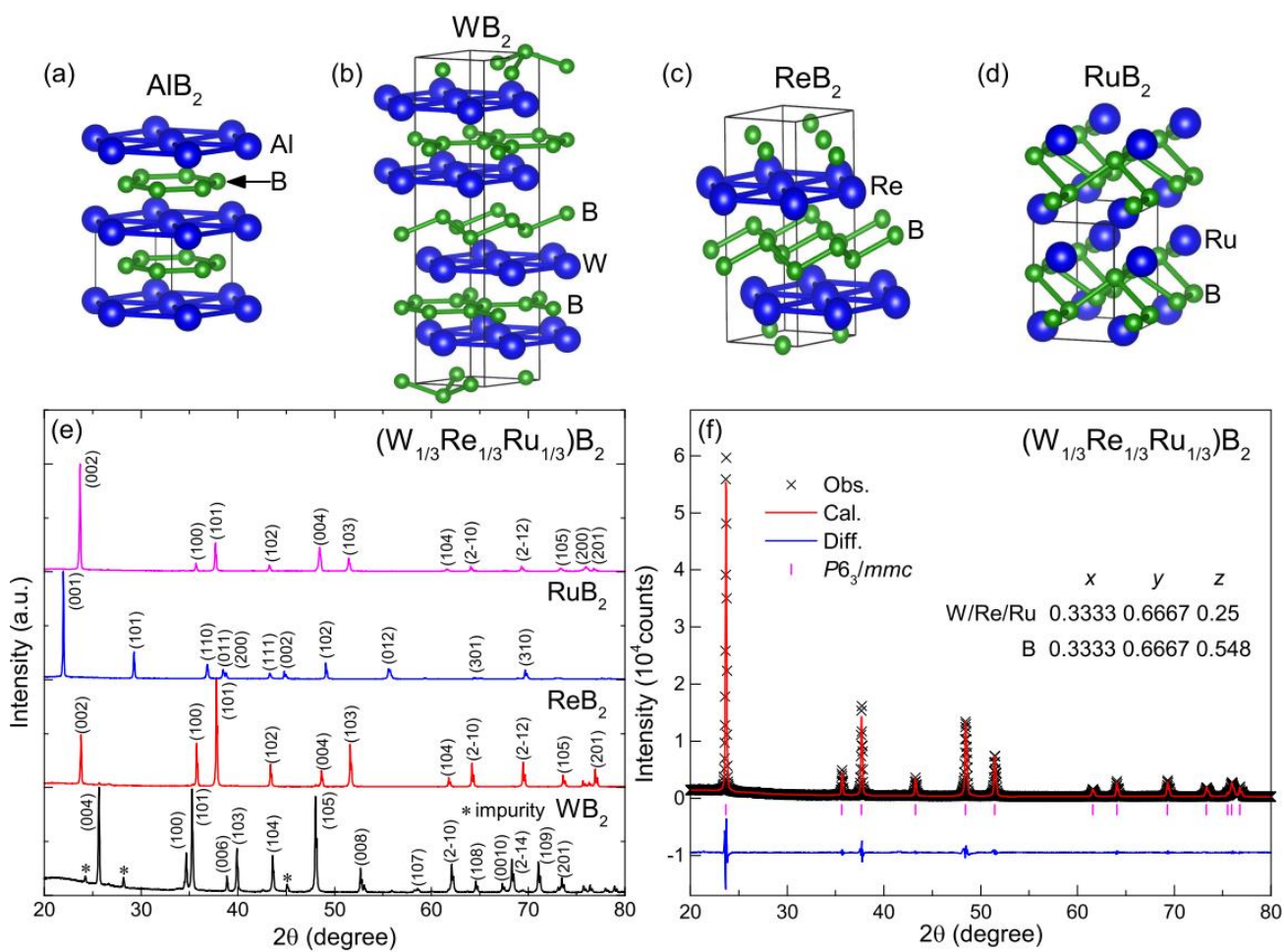

**Fig 1**. (a-d) Schematic structures for $AlB_2$, $WB_2$, $ReB_2$, and $RuB_2$, respectively. (e) XRD patterns for the $(W_{1/3}Re_{1/3}Ru_{1/3})B_2$ MEDB and binary diborides $WB_2$, $ReB_2$ and $RuB_2$. The diffraction peaks are indexed for all samples and those from $WB_4$ impurity are marked by the asterisks. (f) Structural refinement profile for the $(W_{1/3}Re_{1/3}Ru_{1/3})B_2$ MEDB.

**Figure 2**

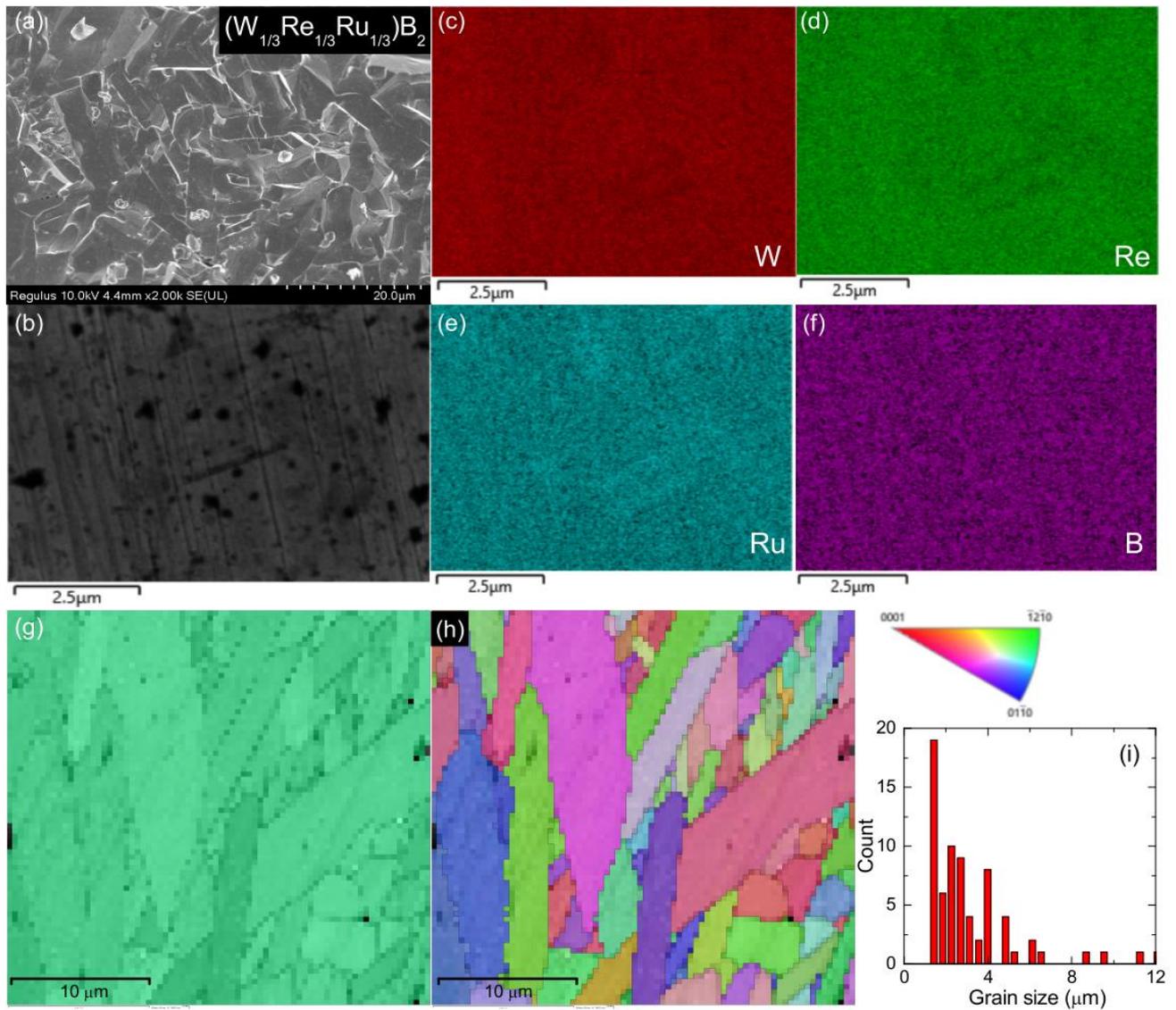

**Fig 2**. (a) SEM image of the a fracture surface of a broken $(W_{1/3}Re_{1/3}Ru_{1/3})B_2$ MEDB with a scale bar of 20 $\mu$m. (b-f) SEM image and corresponding elemental maps of the smoothed surface of the MEDB with a scale bar of 2.5 $\mu$m. (g,h) EBSD phase map and orientation map of the polished surface of the MEDB with a scale bar of 10 $\mu$m. (i) Grain size distribution from panel (h).

**Figure 3**

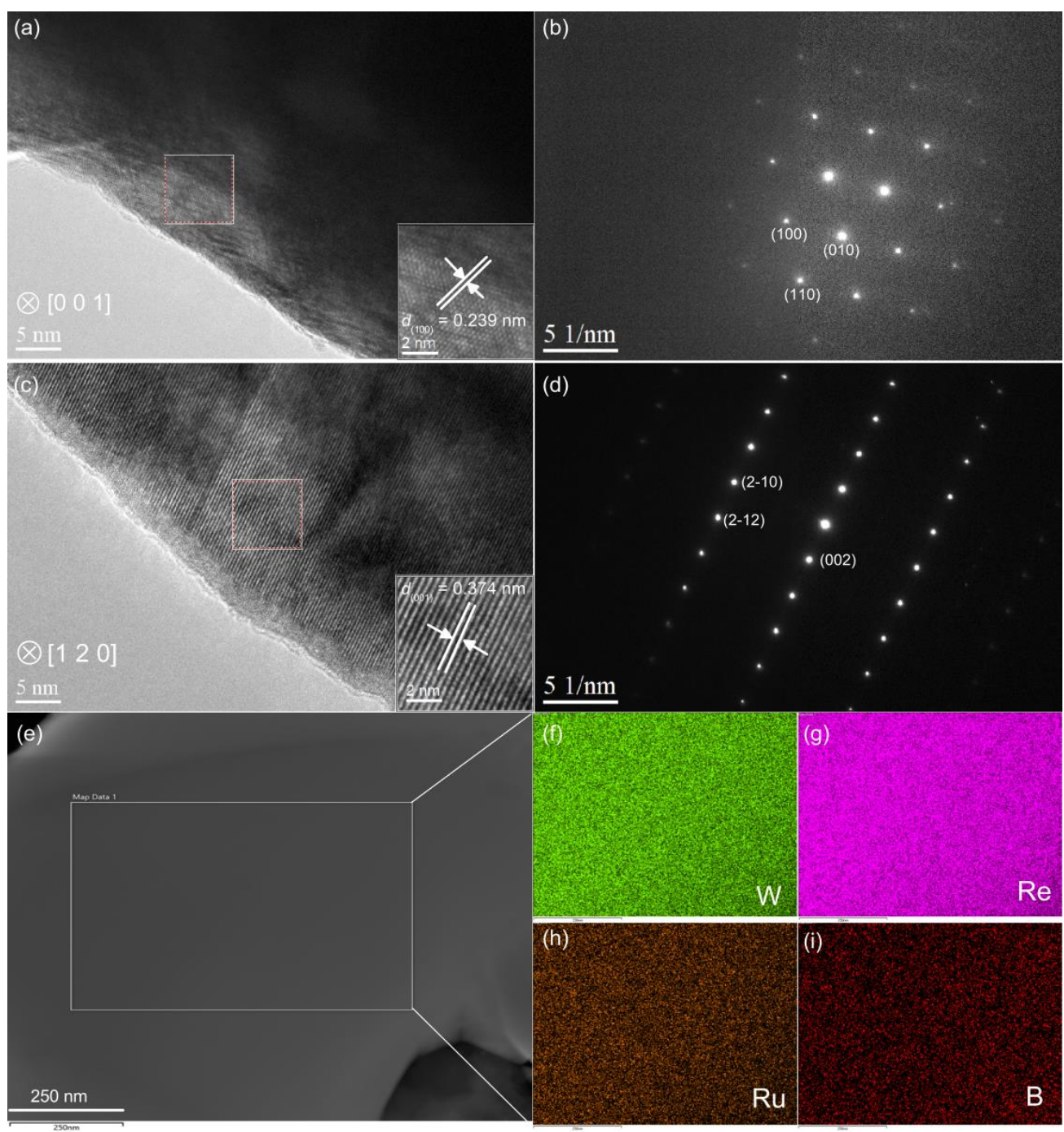

**Fig 3**. (a,b) TEM image, with a scale bar of 5 nm, and corresponding SAED pattern for the $(W_{1/3}Re_{1/3}Ru_{1/3})B_2$ MEDB taken along the [0 0 1] zone axis. In panel (a), the inset shows a zoom of the TEM image and the lattice spacing is indicated. (c,d) Same set of data for the MEDB taken along the [1 2 0] zone axis. In panel (c), the inset shows a zoom of the TEM image and the lattice spacing is indicated. (e) TEM image with a scale bar of 250 nm for the MEDB. (f-i) EDX elemental maps for W, Re, Ru and B, respectively, of the area marked by the frame in panel (e).

**Figure 4**

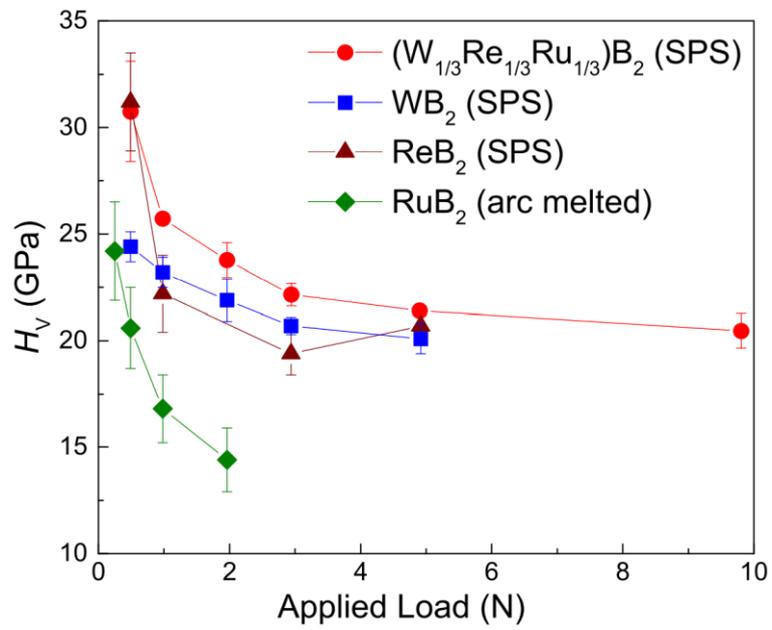

**Fig 4**. Dependence of the Vickers hardness on the applied load for the $(W_{1/3}Re_{1/3}Ru_{1/3})B_2$ MEDB. For comparison, the data sets for binary diborides $WB_2$ [20], $ReB_2$ [21] and $RuB_2$ [22] are also included.

# Supplementary Material for

# "Spark plasma sintering synthesis of ReB$_2$-type medium-entropy diboride (W$_{1/3}$Re$_{1/3}$Ru$_{1/3}$)B$_2$ with high hardness"

Wuzhang Yang[1,2,3], Guorui Xiao[1,2,4], Zhi Ren[1,2,*]

[1] *School of Science, Westlake University, 18 Shilongshan Road, Hangzhou 310024, P. R. China*

[2] *Institute of Natural Sciences, Westlake Institute for Advanced Study, 18 Shilongshan Road, Hangzhou 310024, P. R. China*

[3] *Department of Physics, Fudan University, Shanghai 200433, P. R. China*

[4] *Department of Physics, Zhejiang University, Hangzhou 310027, P. R. China*

\* Corresponding author.

E-mail: renzhi@westlake.edu.cn

**S1. Photograph of the sintered MEDB after surface polishing**

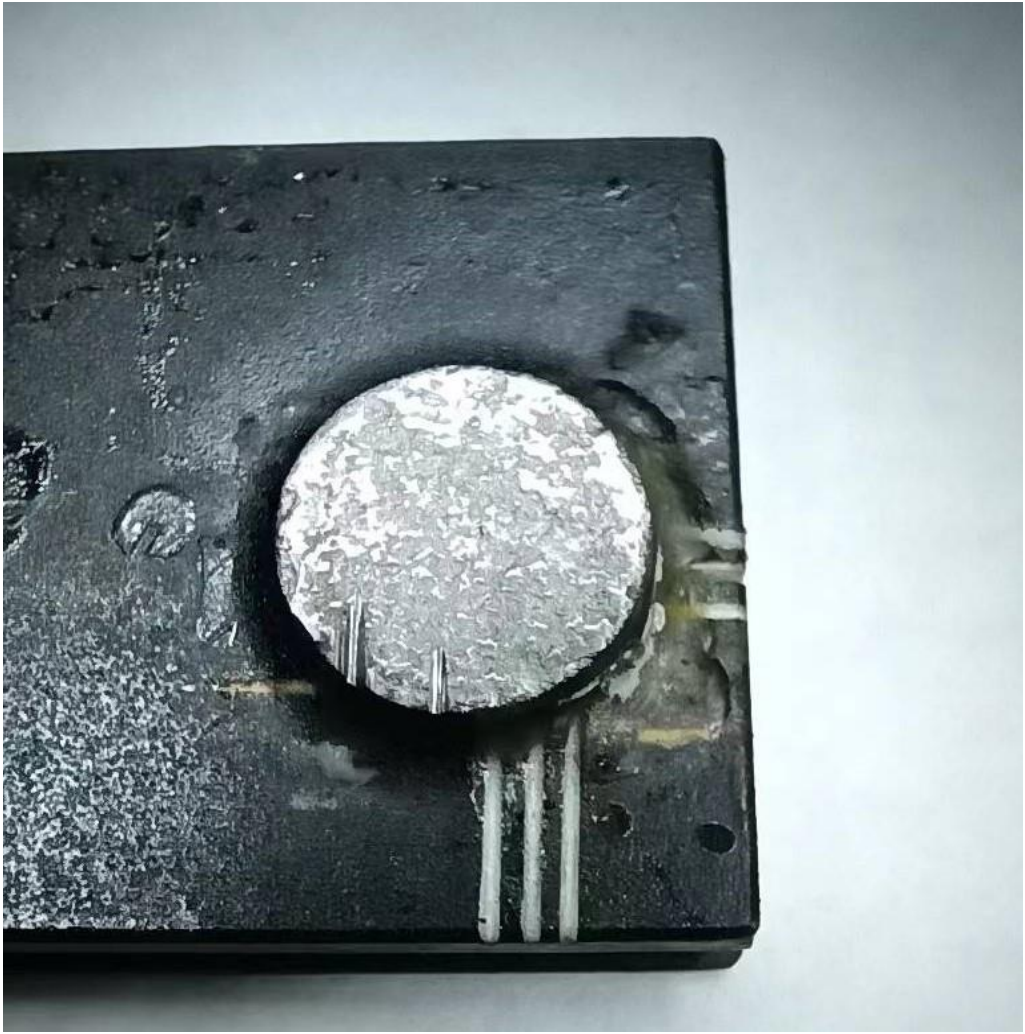

**Fig S1**. Photograph of the SPS sintered $(W_{1/3}Re_{1/3}Ru_{1/3})B_2$ MEDB after surface polishing. The sample has a disk shape with a radius of 10.0 mm and a height of 2.2 mm.

**S2. Indentation images for loads ranging from 0.49 N to 9.8 N**

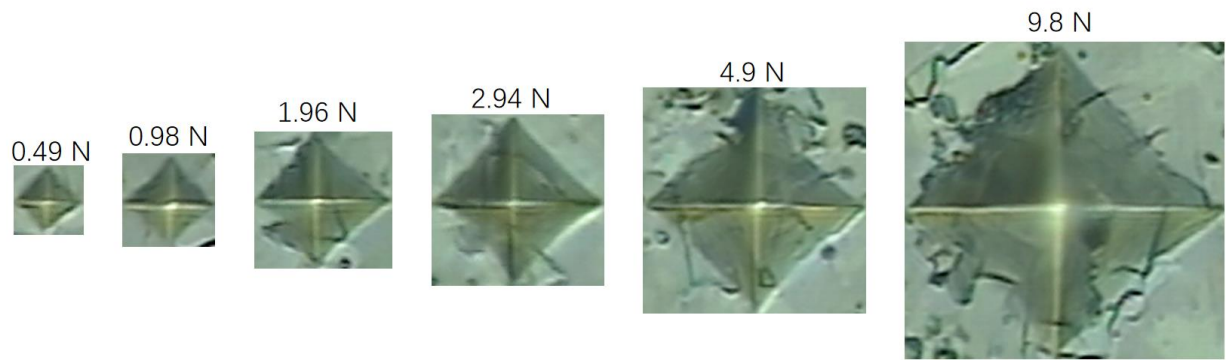

**Fig S2**. Vickers indentation images of the $(W_{1/3}Re_{1/3}Ru_{1/3})B_2$ MEDB for loads varying from 0.49 N to 9.8 N.